\documentclass[a4paper]{jpconf}
\usepackage{graphicx}
\usepackage[table,xcdraw]{xcolor}

\usepackage{hyperref}
\hypersetup{
    colorlinks=true,
    linkcolor=blue,
    filecolor=magenta,      
    urlcolor=blue,
    citecolor=red
}

\begin{document}
\title{\large Blended learning: A data-literate science teacher 
is a better teacher}

\author{J Hanč$^{1}$, D Borovský$^{1}$ and Martina Hančová$^{2}$}

\address{$^1$Institute of Physics, Faculty of Science, Pavol Jozef Šafárik University in Košice, Slovakia}
\address{$^1$Institute of Mathematics, Faculty of Science, Pavol Jozef Šafárik University in Košice, Slovakia}

\ead{jozef.hanc@upjs.sk$^1$}

\begin{abstract}
The COVID-19 pandemic has underscored the importance of blended learning in contemporary physics and, more generally, STEM education. In this contribution, we summarize current pedagogical models of blended learning, such as rotational and flexible non-rotational models, and customizable configurations of physical and virtual learning spaces. With the inevitable integration of digital technology as one of the pillars of blended learning, teachers find themselves in an unprecedented position to not only obtain data more frequently but also analyze it and adjust instruction accordingly. Consequently, we discuss a crucial element of blended learning effectiveness: data management and usage. In this context, data literacy for teaching emerges as an essential skill for effective blended learning, encompassing the ability to transform various data types into actionable instructional knowledge and practices. In other words, current research in physics education shows that a data-literate science teacher is a more prosperous and effective teacher.
\end{abstract}

\section{Introduction}
In spring 2020, the world was shaken by the COVID-19 pandemic, which drastically altered almost every sector of human activity. Businesses, theaters, schools, and entire states experienced lockdowns. This period saw significant losses of human life, overburdened hospitals, economic hardships, and setbacks in sectors like tourism and culture. However, there were silver linings. The crisis highlighted the importance of telemedicine, video banking, and working from home (known as "home office"). Remarkably, we achieved such digital progress in just a few months that would typically span 3-5 years.

\noindent Education was profoundly affected. A vast majority of educational institutions were caught off-guard by the sudden shift to online learning. However, some teachers, classes, and even schools managed to switch to distance online education seamlessly overnight with remarkable agility, proving that students could effectively acquire knowledge online. A common characteristic among these schools \cite{divjakFlippedClassroomsHigher2022}  was their prior successful adoption of either some blended learning instructional model \cite{tuckerBlendedLearningAction2017, tuckerCompleteGuideBlended2022} or flipped learning model \cite{talbertFlippedLearningGuide2017a, bredowFlipNotFlip2021a}, both of which offer a well-balanced and pedagogically sound integration of onsite and online education. The COVID-19 pandemic not only underscored the importance of blended learning in education, including contemporary physics and, more generally, STEM education but also advanced blended learning itself and intensified pedagogical research in this field.

\noindent The aim of our article is to briefly summarize the current status and progress in blended learning (BL), focusing on its key cornerstones, various forms of pedagogical models, and assessment strategies.

\section{Blended Learning Cornerstones: Learning Spaces \& Technology}

Many didactic textbooks, and articles on modern education, but also teacher training courses, pay much less attention to the learning environment than they do to the actual educational forms, methods, means, or technologies. In the context of BL, the learning space becomes a pivotal cornerstone \cite{tuckerBlendedLearningAction2017, basyeGetActiveReimagining2015}. The seamless integration of the physical and digital space into a so-called hybrid learning space adds a new dimension to the perception of where and how students should be educated. In the research literature focused on BL, we found a particularly insightful view of the modern hybrid learning space proposed by the Italian researchers Bocconi and Trentin.

\begin{figure}[h]
    \centering
    \includegraphics[width=0.9\textwidth]{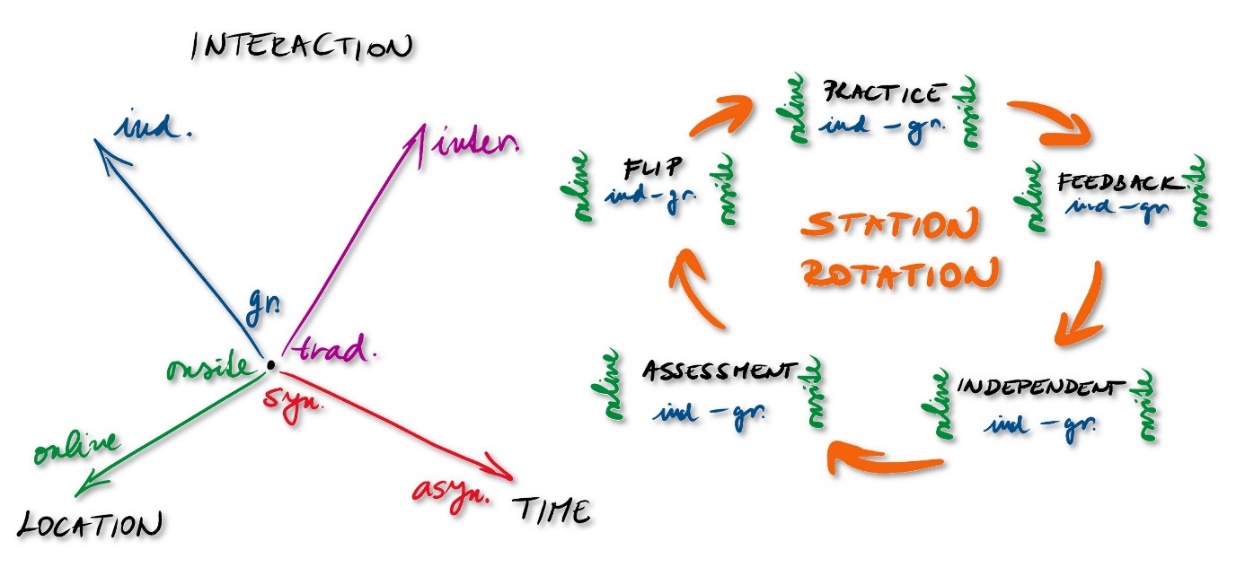}
    \caption{\label{fig1}Learning space dimensions (on the left) and types of stations (on the right) in BL}
\end{figure}

\noindent Based on their perspective \cite{bocconiModellingBlendedSolutions2014}, we can consider several natural, basic criteria, called \textit{dimensions} in blended learning, to divide or classify learning spaces: \textit{location, time}, and \textit{interaction} (graphically depicted in Fig.~\ref{fig1} on the left). Based on the location, we talk about space:

\begin{itemize}
    \item \textit{physical} (real) -- onsite
    \item \textit{digital} (virtual) -- online
\end{itemize}

\noindent Examples of a physical learning space are the traditional classroom and the modern "Future Classroom Lab" (FCL), which we will describe briefly in the following text. As for digital spaces, the landscape spans from standard Learning Management Systems (LMS) like Moodle (\url{https://moodle.org}) or Canvas (\url{https://www.instructure.com/canvas}) to more encompassing educational app ecosystems such as Google Workspace for Education and Microsoft Teams. 

\noindent From a time dimension, we speak of the learning space in which education takes place:

\begin{itemize}
    \item \textit{synchronous} -- simultaneously, often at approximately the same pace, usually in the classroom
    \item \textit{asynchronous} -- not simultaneously, anytime, at one's own pace, typically online
\end{itemize}

\begin{figure}[h]
    \centering
    \includegraphics[width=0.9\textwidth]{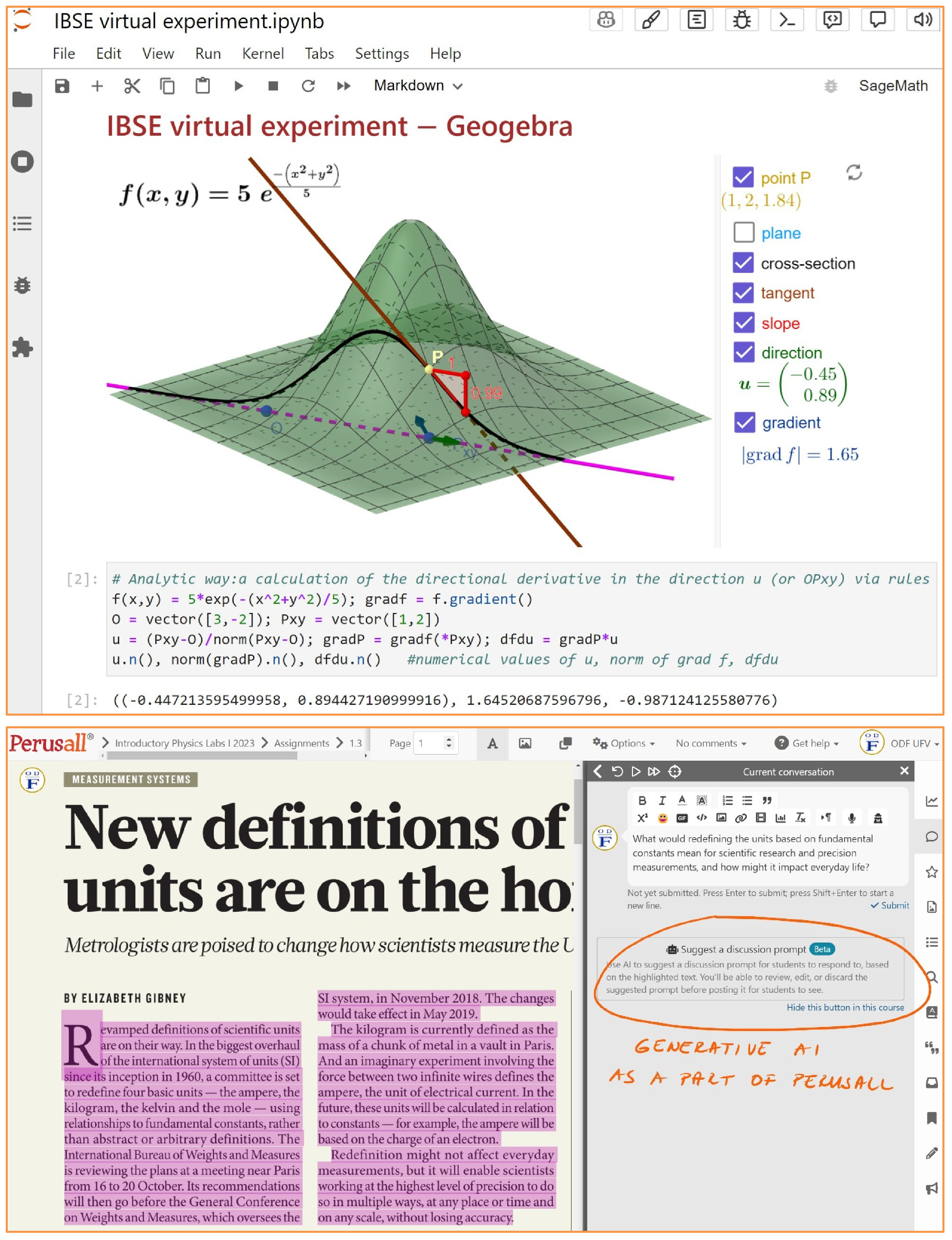}
    \caption{\label{fig2}Two special educational examples of digital learning spaces – the Jupyter Notebook environment (on the upper half) and the social reader Perusall environment (on the lower half).}
\end{figure}

\noindent Especially notable digital spaces in STEM education are two widely used and free collaborative digital platforms.  The open science Jupyter notebooks (\url{https://jupyter.org}) allow effective interactive STEM education \cite{borovskyScientificComputingOpen2023a, gajdosInteractiveJupyterNotebooks2022b, hancOpenPythonbasedDigital2022} where students can collaboratively (synchronously or asynchronously), e.g. interactive visualizations, modeling, programming, and solve genuine, multifaceted interdisciplinary challenging problems (see Fig.~\ref{fig2}, upper half). The social e-reader Perusall (\url{https://www.perusall.com}), one of the key BL tools during the pandemic, facilitates both synchronous but predominantly asynchronous social collaborative viewing, studying, and learning through individual annotations of any digital content \cite{hancSocialReaderPerusall2023a}. 

\noindent {\small\textit{Remark.} Intriguingly, both these platforms currently harness the power of current generative AI tools. In the Jupyter notebooks, a chatbot helps students with coding, clarifications, and scientific computations. With Perusall, AI assists educators in crafting thought-provoking discussion prompts, as depicted in Fig.~\ref{fig2} (the lower half) where the system suggests a relevant question for students based on highlighted content.}


\noindent When considering pedagogical interaction between participants, from the viewpoint of the participant number or the form of teaching, we talk about:

\begin{itemize}
    \item \textit{individual} learning space, usually managed by the student working alone
    \item \textit{group} learning space, most often guided by a teacher, with students working in groups
\end{itemize}

\noindent In terms of interaction, we can also consider the student's activity or the teaching method. This determines whether we distinguish the educational space as:

\begin{itemize}
    \item \textit{traditional} (passive), where traditional teaching methods usually lead to student passivity
    \item \textit{interactive} (active), where it is easy and effective to apply active learning and interactive methods
\end{itemize}

\begin{figure}[h]
    \centering
    \includegraphics[width=0.6\textwidth]{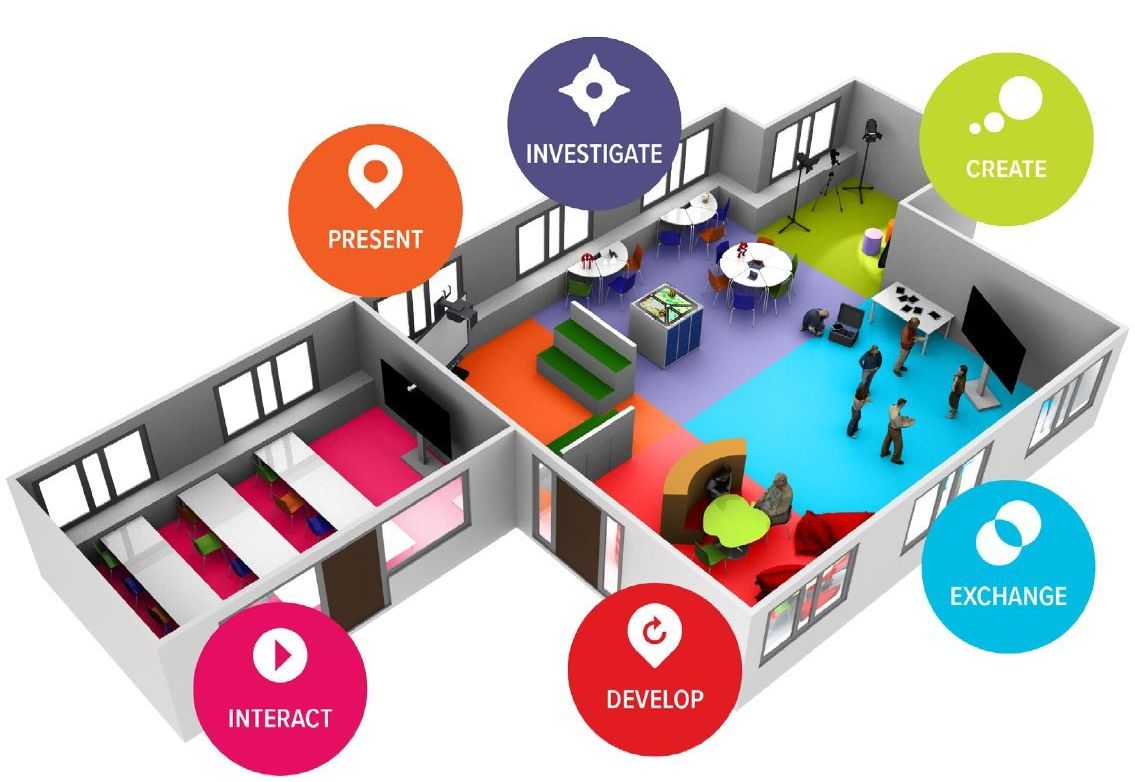}
    \caption{\label{fig3}Learning zones according to FCL in blended learning (image source: \cite{attewellBuildingLearningLabs2019})}
\end{figure}

\noindent One of the innovative active learning spaces is the \textit{Future Classroom Lab} shortly FCL (\cite{attewellBuildingLearningLabs2019}, see Fig.~\ref{fig3}). It was opened in January 2012 (and still improving) in Brussels, within the premises of the European Schoolnet (\url{http://www.eun.org}). 

\noindent The Brussels FCL future classroom consists of six fundamental learning spaces referred to as education zones (Fig.~\ref{fig1}). Their mission is to actively connect, engage, and stimulate students in learning. The specific anticipated didactic use, along with the proposed digital technologies, is described in Table \ref{table1}.

\noindent {\small\textit{Remark.} The fact that the FCL future classroom (\url{https://fcl.eun.org}) was established as part of one of the pilot projects of the European Schoolnet has a simple explanation. European Schoolnet, a non-profit alliance of 32 European education ministries, is dedicated to fostering, creating, and validating innovative educational models that leverage digital technologies. The conception and design of the FCL classroom is a direct result of this commitment, grounded in extensive research. This model classroom not only provides a space where teachers can explore and experiment with advanced pedagogical methods and digital resources but also serves as a beacon, guiding educators and school administrations toward understanding the importance of the learning environment and inspiring them to transform their classrooms into more active learning spaces.} 

\noindent Sometimes, e.g. in the case of the BYOD (Bring Your Own Device) policy \cite{attewellBringYourOwn2017} in BL, it is also useful to consider the 5$^{\mathrm{th}}$ dimension for learning spaces classification with respect to educational settings: 

\begin{itemize}
    \item \textit{formal} (e.g. a regular classroom; a digital space of LMS Moodle) 
    \item \textit{informal} (e.g. “in-between” hallway space or a bus; a digital space of Facebook or YouTube)
\end{itemize}

\noindent Teachers and school leaders can find detailed and practical guidelines on how to imagine and construct hybrid active learning spaces for both formal and informal BL education in \cite{tuckerBlendedLearningAction2017, basyeGetActiveReimagining2015, attewellBringYourOwn2017}. Especially for STEM education, more specific insights and examples are available in \cite{talbertSpaceLearningAnalysis2019}.

\begin{table}[h]
\caption{Learning zones in the Brussels Future Classroom Lab (FCL)}
\label{table1}
\small
\begin{tabular}{|l|l|l|}
\hline
\rowcolor[HTML]{000000} 
{\color[HTML]{FFFFFF} \textbf{zone}} &
  {\color[HTML]{FFFFFF} \textbf{educational activities, methods}} &
  {\color[HTML]{FFFFFF} \textbf{educational (digital) means}} \\ \hline
\rowcolor[HTML]{C0C0C0} 
{\color[HTML]{000000} \textbf{“investigate”}} &
  {\color[HTML]{000000} \begin{tabular}[c]{@{}l@{}}inquiry-based learning, discovery, \\ problem-solving, projects, critical \\ thinking\end{tabular}} &
  {\color[HTML]{000000} \begin{tabular}[c]{@{}l@{}}laptops, mobile devices, measuring probes, \\ digital microscopes, robots, online labs, \\ simulations\end{tabular}} \\ \hline
{\color[HTML]{000000} \textbf{“create”}} &
  {\color[HTML]{000000} \begin{tabular}[c]{@{}l@{}}creation of (non)digital content, \\ gamification, intellectual property, \\ creativity\end{tabular}} &
  {\color[HTML]{000000} \begin{tabular}[c]{@{}l@{}}audiovisual equipment, digital content \\ creation software, green screen \\ technology, printers (including 3D)\end{tabular}} \\ \hline
\rowcolor[HTML]{C0C0C0} 
{\color[HTML]{000000} \textbf{“exchange”}} &
  {\color[HTML]{000000} \begin{tabular}[c]{@{}l@{}}collaboration, teamwork, peer \& group \\ discussion, mind mapping, \\ brainstorming, small groups\end{tabular}} &
  {\color[HTML]{000000} \begin{tabular}[c]{@{}l@{}}interactive whiteboards, interactive \\ projectors, electronic flipcharts, mind\\  mapping software\end{tabular}} \\ \hline
{\color[HTML]{000000} \textbf{“develop”}} &
  {\color[HTML]{000000} \begin{tabular}[c]{@{}l@{}}informal individual learning, online \\ education, personalization\end{tabular}} &
  {\color[HTML]{000000} \begin{tabular}[c]{@{}l@{}}laptops, mobile devices, audio players,\\  educational games, videos, documents\end{tabular}} \\ \hline
\rowcolor[HTML]{C0C0C0} 
{\color[HTML]{000000} \textbf{“interact”}} &
  {\color[HTML]{000000} \begin{tabular}[c]{@{}l@{}}peer discussion \& communication, peer \\ learning, individual learning, critical \\ thinking\end{tabular}} &
  {\color[HTML]{000000} \begin{tabular}[c]{@{}l@{}}laptops, mobile devices, e-voting, adaptive \\ or interactive educational software, HD \\ screen\end{tabular}} \\ \hline
{\color[HTML]{000000} \textbf{“present”}} &
  {\color[HTML]{000000} \begin{tabular}[c]{@{}l@{}}ability to present, explain, share, \\ communicate\end{tabular}} &
  {\color[HTML]{000000} \begin{tabular}[c]{@{}l@{}}HD screen, projector, presentation tools, \\ online publishing\end{tabular}} \\ \hline
\end{tabular}
\end{table}

\section{Blended Learning in Action: Pedagogical Models \& Configurations}

The above classification of learning spaces allows us to present the important theoretical frameworks connected to BL. The very natural fundamental pedagogical principle of BL, from which all successful and efficient BL pedagogical models emerge and strive to adhere, states \cite{tuckerBlendedLearningAction2017, hancOpenPythonbasedDigital2022}:

\begin{quote}
    \textit{Let the students do \textbf{easier things} alone, usually independently in their individual space, online, asynchronously, and \textbf{harder, more demanding, challenging things} together with the teacher as a guide in a common group space, usually onsite (face-to-face), synchronously.}
\end{quote}

\noindent The complexity of educational goals, activities, tasks, and required cognitive processes is determined not only by the direct experience of the teacher but primarily by the second theoretical framework related to this principle, the Revised Bloom's Taxonomy (2001). According to \cite{tuckerBlendedLearningAction2017, talbertFlippedLearningGuide2017a, ramirezInClassFlipStudentCentered2022}, BL provides a wide plethora of instructional or pedagogical models set up on flexible and adaptable pedagogical configurations of physical and virtual online learning spaces, allowing highly effective and successful education under the most diverse conditions of school practice. Over the last decade, through innovative designs and proven practices, we can categorize BL models into two main groups \cite{tuckerBlendedLearningAction2017, ramirezInClassFlipStudentCentered2022}:

\begin{itemize}
    \item \textit{rotation} models which can be, e.g., a BL (or in-class flip) station work configured as sequential, mixed, looped, or balanced, but it also includes the standard out-of-class flipped learning
    \item \textit{flexible (‘flex’) non-rotation} models which can have, e.g., a form of solo, duo, or group work
\end{itemize}

\noindent The most pedagogically significant aspect of these BL models is that they not only allow for the application of the flipped learning principles but also effortlessly incorporate every interactive teaching method such as inquiry-based education \cite{constantinouWhatInquiryBasedScience2018}, team and project-based learning \cite{jonesFlippedProjectBased2018}, or peer instruction \cite{mazurJustinTimeTeachingPeer2009, millerUseSocialAnnotation2018a}. 

\noindent \textit{Remark.} Regarding the well-known \textit{Peer Instruction} method, it is worth mentioning that the digitalization prompted by pandemic restrictions led to an asynchronous version of the method. In Perusall, a teacher can pose a question related to a specific part of the study material (as shown in Fig.~\ref{fig2}, lower half, created with the help of AI). This question can be multiple-choice or open and students can respond to it asynchronously. The social interaction can be set, using a special button “puzzle piece”, so that students cannot see the answers and comments of others until they provide their own response.

\noindent To gain a clearer understanding, let's delve deeper into one of these BL instructional models that we are familiar with and frequently utilize in our educational practice.

\begin{table}[]
\caption{Type of stations in blended learning rotation models}
\label{table2}
\small
\begin{tabular}{|l|l|l|}
\hline
\rowcolor[HTML]{000000} 
{\color[HTML]{FFFFFF} \textbf{station}} &
  {\color[HTML]{FFFFFF} \textbf{educational activities, methods}} &
  {\color[HTML]{FFFFFF} \textbf{educational means}} \\ \hline
\rowcolor[HTML]{C0C0C0} 
{\color[HTML]{000000} \textbf{flip station}} &
  {\color[HTML]{000000} \begin{tabular}[c]{@{}l@{}}self-paced learning, multimedia content \\ review, direct instructions \& guidelines, \\ mini-lectures\end{tabular}} &
  {\color[HTML]{000000} \begin{tabular}[c]{@{}l@{}}videos, slides, flashcards, simulations, \\ online sources, reading materials, student-\\ made materials, podcasts\end{tabular}} \\ \hline
{\color[HTML]{000000} \textbf{\begin{tabular}[c]{@{}l@{}}practice \\ station\end{tabular}}} &
  {\color[HTML]{000000} \begin{tabular}[c]{@{}l@{}}active learning, inquiry-based activities, \\ mastery of content, application of \\ Bloom's taxonomy, scaffolding\end{tabular}} &
  {\color[HTML]{000000} \begin{tabular}[c]{@{}l@{}}worksheets, online tasks, games, \\ discussions, speaking activities, pair work, \\ group creation, writing, experiments, arts \\ \& crafts, videos, recordings, role-plays\end{tabular}} \\ \hline
\rowcolor[HTML]{C0C0C0} 
{\color[HTML]{000000} \textbf{\begin{tabular}[c]{@{}l@{}}independent \\ station\end{tabular}}} &
  {\color[HTML]{000000} continuous learning, individual work} &
  {\color[HTML]{000000} \begin{tabular}[c]{@{}l@{}}reviews, hands-on activities, games, crafts, \\ worksheets, online activities, silent reading\end{tabular}} \\ \hline
{\color[HTML]{000000} \textbf{\begin{tabular}[c]{@{}l@{}}teacher \\ support station\end{tabular}}} &
  {\color[HTML]{000000} \begin{tabular}[c]{@{}l@{}}clarification of questions, tutoring, \\ teacher feedback, mini-lectures\end{tabular}} &
  {\color[HTML]{000000} \begin{tabular}[c]{@{}l@{}}direct interaction with the teacher for \\ content clarification, tutoring, and \\ feedback\end{tabular}} \\ \hline
\rowcolor[HTML]{C0C0C0} 
{\color[HTML]{000000} \textbf{\begin{tabular}[c]{@{}l@{}}feedback \\ station\end{tabular}}} &
  {\color[HTML]{000000} \begin{tabular}[c]{@{}l@{}}peer feedback, learning feedback, lesson \\ feedback\end{tabular}} &
  {\color[HTML]{000000} \begin{tabular}[c]{@{}l@{}}online tools, post-its, worksheets, videos, \\ written tasks\end{tabular}} \\ \hline
{\color[HTML]{000000} \textbf{\begin{tabular}[c]{@{}l@{}}assessment \\ station\end{tabular}}} &
  {\color[HTML]{000000} \begin{tabular}[c]{@{}l@{}}assessment of content mastery through \\ quizzes, tests, or cumulative knowledge-\\ based activities\end{tabular}} &
  {\color[HTML]{000000} \begin{tabular}[c]{@{}l@{}}quizzes, tests, cumulative knowledge-\\ based activities\end{tabular}} \\ \hline
\end{tabular}
\end{table}

\noindent This model is rotational using out-of-class flip and mixed station rotation (schematically shown in Fig.~\ref{fig1}, on the right). Central to the station rotation models is the concept of a \textit{station} as a learning space. For instance, learning zones in the FCL can be viewed as distinct stations. Table \ref{table2} offers a comprehensive pedagogical description and classification of stations frequently used in BL rotation models.

\begin{figure}[h]
   \centering 
   \includegraphics[width=0.9\textwidth]{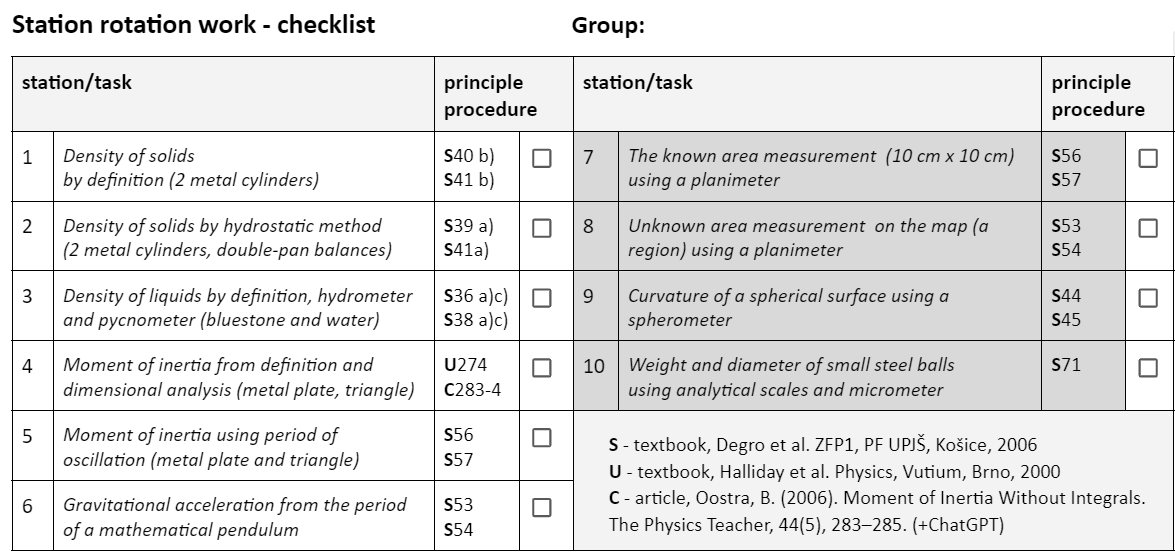}
    \caption{\label{fig4} An example of a station rotation checklist.}
\end{figure}

\noindent In one of our university courses, Physics Practical I -- introductory physics labs, last academic year 2022/2023, we had 6 groups (typically 2-3 students in each one) and 11 lab tables, stations -- 1 teacher support station, 6 practice stations, and 4 stations which can serve both as flipped or independent (see Fig.~\ref{fig4}). In accordance with the out-of-class flip approach, before the face-to-face interaction students asynchronously in their individual spaces (often home), but together collaboratively study the topic from a digital source in Perusall. During this study phase, students highlight challenging sections of the text, engage in collective discussions, clarify concepts for one another, answer questions posed by peers and/or the teacher, and show agreement with others by using a "+1" vote.


\noindent Then during two consecutive lab sessions (each lasting 135 minutes), groups rotate each 20-25 minutes between lab tasks at stations. If some group finish before the next station is available, to avoid a “station traffic jam“ the group can choose a work at flip station where can study new material dealing with the task If no students seek assistance at the teacher station, the teacher becomes a guide. To maintain organization and prevent confusion, students receive a checklist (Fig.~\ref{fig4}) that helps them track and work at the stations they should visit. The results indicate that such intensive retrieval with interleaved practice significantly enhances students' experimental skills. Specifically, students mastered measurements using vernier calipers, micrometers, double-pan balances, digital scales, and setting up simple experimental apparatuses. More educational examples from our blended learning can be found in \cite{gajdosInteractiveJupyterNotebooks2022b, hancOpenPythonbasedDigital2022, hancSocialReaderPerusall2023a, pankovaPracticalStrategiesFormative2017}.

\section{Blended Learning Effectiveness: Assessment \& Data Management}

One of the greatest benefits of blended learning is the easy implementation of innovative and effective student assessment methods such as \textit{assessment of/for/as learning} \cite{hancSocialReaderPerusall2023a, earlAssessmentLearningUsing2013} or \textit{mastery grading} \cite{bergmannMasteryLearningHandbook2022a}. Using these methods through technology, teachers find themselves in an unprecedented position to obtain and analyze educational data more frequently and in larger volumes. Such data can be transformed into actionable instructional insights, leading to more effective teaching \cite{tuckerBlendedLearningAction2017}.

\noindent As the research has also clearly demonstrated that effective data usage in education is among the common and key characteristics of high-performing schools with outstanding student results \cite{mandinachWhatDoesIt2016, fildermanDataLiteracyTraining2022}, one of the newest competencies of an effective teacher is the teacher’s data literacy. According to \cite{mandinachWhatDoesIt2016}, data literacy for teaching is the ability to transform information into actionable instructional knowledge and practices by collecting, analyzing, and interpreting all types of data (assessment, school climate, behavioral, snapshot, longitudinal, moment-to-moment, etc.) to help determine instructional steps.

\noindent Johnson and Christensen \cite{johnsonEducationalResearchQuantitative2016} emphasize the importance of teachers acting as action researchers, armed with essential methodological knowledge and data literacy, which enhances pedagogical practices and boosts job satisfaction. To support teachers to be data-literate in the effective use of data, training is necessary. However, based on our experience and that of others \cite{mandinachWhatDoesIt2016, fildermanDataLiteracyTraining2022}, offering such training and convincing in-service teachers to participate is quite challenging due to the demanding nature of teaching and the significant workload.

\section{Future perspectives in BL data management}

Today, in our current scientific projects mentioned in the acknowledgments, we are intensively exploring the use of modern open data science tools to simplify the handling of educational data. From what we have found so far, these tools greatly streamline the process with just a few clicks, eliminating the need for extensive setups or substantial software investments (see our papers \cite{borovskyScientificComputingOpen2023a, gajdosInteractiveJupyterNotebooks2022b, hancOpenPythonbasedDigital2022, hancSocialReaderPerusall2023a}). Furthermore, with the rise of generative AI, such as ChatGPT, these tools have become more accessible to BL teachers and researchers without programming skills. This accessibility will allow a broader range of teachers and educators to enhance their teaching methods and instruction dealing with BL. It is also important to note that these teachers can provide us as researchers with educational measurement data, which, when analyzed using new statistical methods, can reveal fresh insights and findings for physics and STEM education.

\section{References}
\bibliographystyle{iopart-num}
\bibliography{DIDFYZ2023_ref}

\section*{Acknowledgments}
This work is supported by the Slovak Research and Development Agency under the Contract no. APVV-22-0515, APVV-21-0216 and APVV-21-0369.

\end{document}